\documentclass{article}
\usepackage{ifpdf}
\usepackage{cite}
\usepackage{amsmath}
\usepackage{array}
\usepackage{fixltx2e}
\usepackage{stfloats}
\usepackage{url}

\usepackage{graphicx,psfrag,epsf}
\usepackage{mathtools}
\usepackage{booktabs}
\usepackage[english]{babel} 
\usepackage{amsmath,amsfonts,amsthm}
\usepackage{chngcntr}
\usepackage{ amssymb }
\usepackage{array}
\usepackage{booktabs} 
\usepackage{setspace}
\usepackage{fix-cm}
\usepackage{xcolor}
\usepackage[normalem]{ulem}
\usepackage{hyperref}
\hypersetup{colorlinks,urlcolor=blue}
\usepackage{url}
\usepackage{tikz}
\usepackage{cool}
\usepackage{enumitem}
\usepackage{soul}
\usepackage{threeparttable}
\usepackage{tablefootnote}
\usepackage{color}
\usepackage{tabularx}
\usepackage{footnote}
\usepackage{algorithm}
\usepackage[noend]{algpseudocode}

\makeatletter
\DeclareUrlCommand\ULurl@@{%
  \def\UrlLeft{\bgroup}%
  \def\UrlRight{\egroup}}
\def\ULurl@#1{\hyper@linkurl{\ULurl@@{#1}}{#1}}
\DeclareRobustCommand*\ULurl{\hyper@normalise\ULurl@}

\newcommand{\distas}[1]{\mathbin{\overset{#1}{\kern\z@\sim}}}%
\newcommand{\bm}[1]{\mathbf{#1}}
\newsavebox{\mybox}\newsavebox{\mysim}
\newcommand{\distras}[1]{%
  \savebox{\mybox}{\hbox{\kern3pt$\scriptstyle#1$\kern3pt}}%
  \savebox{\mysim}{\hbox{$\sim$}}%
  \mathbin{\overset{#1}{\kern\z@\resizebox{\wd\mybox}{\ht\mysim}{$\sim$}}}%
}
\DeclareMathOperator{\nullop}{Null}
\newtheorem{theorem}{Theorem}
\newtheorem{remark}{Remark}

\newtheorem{lemma}[theorem]{Lemma}
\newtheorem{corollary}{Corollary}
\newtheorem{proposition}{Proposition}
\newtheorem{assumption}{Assumption}
\newcolumntype{C}[1]{>{\centering\let\newline\\\arraybackslash\hspace{0pt}}m{#1}}

\newcommand{\be}{\begin{equation}}
\newcommand{\ee}{\end{equation}}
\newcommand{\bi}{\begin{itemize}}
\newcommand{\ei}{\end{itemize}}
\newcommand{\ben}{\begin{enumerate}}
\newcommand{\een}{\end{enumerate}}

\newcommand{\R}{\mathbb{R}}

\allowdisplaybreaks
\DeclareMathOperator*{\argmin}{\arg\!\min}

\DeclareMathOperator*{\rank}{rank}
\makeatother
\usepackage[a4paper,left=1in,right=1in,top=1in,bottom=1in]{geometry}
\let\oldbibliography\thebibliography
\renewcommand{\thebibliography}[1]{\oldbibliography{#1}\setlength{\itemsep}{0pt}}

\title{A Graph-Prediction-Based Approach for Debiasing Underreported Data}
\author{Hanyang Jiang, 
        Yao Xie
\thanks{Hanyang Jiang (e-mail: \ULurl{scottjhy@gatech.edu}) and Yao Xie (e-mail: \ULurl{yao.xie@isye.gatech.edu}) are with the H. Milton Stewart School of Industrial and Systems Engineering, Georgia Institute of Technology, Atlanta, GA.}
}
\date{\today}

\begin{document}
\maketitle
\begin{abstract}
We present a novel Graph-based debiasing Algorithm for Underreported Data (GRAUD) aiming at an efficient joint estimation of event counts and discovery probabilities across spatial or graphical structures. This innovative method provides a solution to problems seen in fields such as policing data and COVID-$19$ data analysis. Our approach avoids the need for strong priors typically associated with Bayesian frameworks. By leveraging the graph structures on unknown variables $n$ and $p$, our method debiases the under-report data and estimates the discovery probability at the same time. We validate the effectiveness of our method through simulation experiments and illustrate its practicality in one real-world application: police 911 calls-to-service data.
\end{abstract}


%

\vspace{-0.1in}
\section{Introduction}
\label{intro}

Bias in data collection is a prevalent issue in many real-world applications due to a variety of reasons. One common scenario is under-reporting, as elucidated by \cite{hazell2006under}. For example, police 911 calls-to-service reports, as\cite{watson2015estimating} illustrates, potentially omit a significant number of unrecorded incidents. Similarly, during the COVID-19 pandemic, data collection, as indicated by \cite{shuja2021covid}, only accounted for those individuals who tested positive, which overlooked asymptomatic individuals and those who hadn't undergone testing, further perpetuating data bias.


The challenge in addressing underreporting data is that there exists a substantial identifiability issue. For instance, while the count of observed cases, denoted by $y$, is known, the count of unobserved instances is not uniquely determined. This is due to the fact that there are infinitely many solutions for the equation $y=np$ when only $y$ is known. 
The problem of estimating the probability $p$ in a Binomial($n$, $p$) distribution when the number of trials $n$ is known has been thoroughly addressed in the classic statistical literature. However, the circumstance where both $n$ and $p$ are unknown is much harder and more interesting. This gives rise to the binomial $n$ problem \cite{dasgupta2005estimation}. In the realm of statistics, this problem is a well-known issue when it comes to one-dimensional cases. Traditionally, the approach to resolving it involves utilizing Bayesian methodologies, as detailed in works like \cite{basu2003ch, basu2001bayesian,draper1971bayesian,feldman1968estimation,raftery1987inference}. However, this leads to the secondary challenge of selecting an appropriate prior.

 Following the setting mentioned before, where there are $n$ possible events, each being observed with a probability $p$. We are then confronted with a count, $y$, which is modeled as a Binomial random variable, $\mbox{Binomial}(n, p)$. Our primary focus is on estimating the parameters $n$ and $p$ while relying exclusively on the observations of $y$. A significant complication arises from the fact that the expected value of $y$ is $np$. In order to circumvent the identifiability issue, additional information or assumptions are needed. In many real-world datasets, each observation carries spatial information, potentially enabling us to model the problem with a graph and take advantage of graph properties. It is also possible to include contextual information to restrict the value of $n$ or $p$. This paper presents such a graph-prediction-based approach for debiasing underreported data called GRAUD.

\subsection{Related work}

The initial investigation of concurrently estimating the parameters $n$ and $p$ was undertaken by Whitaker \cite{whitaker1914poisson}, Fisher \cite{fisher1941negative}, and Haldane \cite{haldane1941fitting}. They put forth the Method of Moments Estimators (MMEs) and Maximum Likelihood Estimates (MLEs). Fisher, however, was dismissive of the problem, positing that with an adequately large number of observations $k$, $n$ would become known. This stance, while technically accurate, proves problematic in practice for small $p$ values, as $k$ would need to be excessively large for reliable certainty in $n$.

Classical methods and their asymptotic properties underwent further scrutiny by Olkin \cite{olkin1981comparison}, Carroll and Lombard \cite{carroll1985note}, and Casella \cite{casella1986stabilizing}. They noted that the MME and MLE estimators, introduced by Haldane and Fisher respectively, were exceptionally sensitive to minor fluctuations in the count data, leading to instability. This behavior emerges when the sample mean and variance are closely matched. Olkin introduced two stabilized versions of these estimators to counter this instability: jackknife-stabilized MLE:S and ridge-stabilized MME:S \cite{olkin1981comparison}, which showed improved performance over their standard counterparts.
Carroll and Lombard \cite{carroll1985note} later proposed a different stabilization approach. They recommended a new estimator MB($a,b$) based on the assumption of a beta prior distribution for $p$ and an integrated likelihood maximization. Casella \cite{casella1986stabilizing} further explored scenarios where stabilized estimators outperformed traditional ones.

Recently, DasGupta and Rubin \cite{dasgupta2005estimation} introduced two innovative, more efficient estimators. The first one is a novel moment estimator that utilizes the sample maximum, mean, and variance, while the second one introduces a bias correction for the sample maximum. These estimators have shown superior performance in various scenarios and their asymptotic properties have been thoroughly studied.
The binomial $n$ problem has also been considered from a Bayesian viewpoint by multiple authors. For example, Draper and Guttman \cite{draper1971bayesian} proposed a Bayes point estimate that presumes a discrete uniform distribution for $n$ over a set ${1,2,…,N}$. Other researchers have proposed Bayes estimators based on various prior distributions for $n$ \cite{kahn1987cautionary, hamedani1988bayes, gunel1989estimation}. While Bayesian approaches have successfully mitigated some difficulties associated with classical approaches, they lack grounding in asymptotic theory, thus better suiting "small" practical problems. 
In the scenario where $n$ is to be estimated with a known $p$, Feldman and Fox \cite{feldman1968estimation} have provided estimates based on MLE, MVUE, and MME and explored their asymptotic properties.
Despite these efforts, the binomial $n$ problem remains fundamentally challenging when $p$ is unknown. The problem is characterized by intrinsic instability, and both $n$ and $p$ parameters have been proven not to be unbiasedly estimable \cite{dasgupta2005estimation}, resulting in difficulties in obtaining reliable estimates. The most common issue across estimators is the severe underestimation of $n$, particularly when $n$ is large or $p$ is small. Without replication, drawing inferences about $n$ becomes impossible.

In our work, we approach the binomial n problem from a fresh perspective and setting. Instead of requiring multiple observations to deduce statistics and discern statistical properties, we leverage the inherent graph structure in the problem and approach it from a signal-processing standpoint. While our methodology bears some resemblance to the concept of convex demixing, it does not necessitate any form of sparsity. Our technique is user-friendly and demonstrates a high level of accuracy and efficiency.
The rest of the paper is organized as follows. Section \ref{setting} introduces the problem setting. Section \ref{alg} introduces our proposed graph-based debiasing algorithm called GRAUD. Section \ref{thy} provides a thorough theoretical analysis of GRAUD. Section \ref{exp} demonstrates the effectiveness of our debiasing algorithm in synthetic data and explores its usefulness for solving real-world problems on real emergency ($911$) call data. Section \ref{con} concludes with thoughts on future work.

\section{Problem setting}
\label{setting}
We define the graph $G=(V,E)$, where $|V|=M$ is the vertex set and $E$ is the edge set. Consider the following scenario where a collection of incidents happens at the vertices of $G$. At node $i$, the observed number of incidents is denoted as $y_i$. We further assume that at each node, the probability of observing an incident is $p_i$, and the true number of incidents, though unknown, is $n_i$. One important thing is that there can be some incidents that happen but unobserved.
This problem can be expressed as a binomial model in the form of:
\begin{equation}
y_i \sim \mbox{Binomial}(n_i, p_i),
\end{equation}
and the expected value of $y_i$ is $\mathbb E[y_i] = n_i p_i$ for $i=1,\cdots,M$. Our target is to jointly estimate the set ${p_i, n_i}$ given the observed data ${y_i}$ and the graph structure, which can be denoted by the adjacency matrix $A$. It's worth noting that there is an identifiability issue \cite{dasgupta2005estimation} associated with this problem, and hence, additional structure and regularization need to be imposed to make it meaningful and solvable.

In numerous practical applications, spatial information forms an inherent part of the data \cite{zhu2023generalized}. For example, the police data records the crimes happening across the state. By partitioning a state into various regions and representing each region by a node, a count $y$ of crimes can be assigned to each vertex. We want to estimate the true number of incidents $n_i$ happenning and the discovery probability $p_i$ for each region. Inspired by this setup, we put forth two reasonable assumptions to address the identifiability issue inherent to this problem.

The first assumption is that the discovery probabilities $p_i$ are spatially smooth, which means that the probability $p_i$ should not vary significantly between neighbors. The smoothness can be measured quantitatively and indicate the degree of change in the graph signal w.r.t. the underlying graph \cite{qiu2017time}. This originates from the idea that factors contributing to the discovery or reporting of incidents, such as law enforcement presence or public awareness, are likely to be fairly consistent across adjacent regions. In other words, these factors tend to exhibit a smooth spatial variation rather than abrupt changes, leading to relatively uniform discovery probabilities in neighboring regions. 

The second assumption is that the true count of incidents, represented as $n_i$, is not arbitrary but instead determined by an underlying model with noises. The model is influenced by socioeconomic factors and characteristics of each region, such as population density, average income, education level, and other pertinent demographic or geographic factors. For example, a richer region with a higher average income tends to have better security, which leads to a higher discovery rate. Hence, the true incident count $n_i$ in a particular region can be viewed as a manifestation of these socioeconomic factors.

Next, we proceed to provide mathematical formulations of our assumptions. The
graph Laplacian quadratic form \cite{shuman2013emerging} is often used to represent graph smoothness, we posit that the quantity $p^TLp$ should be small. Here, $p = [p_1, \ldots, p_M]^T\in\mathbb{R}^{M}$ is a vector of discovery probabilities, and $L = D-A \in\R^{M\times M}$ denotes the graph Laplacian, where $D$ stands for the degree matrix. Based on the equation

\begin{equation}
p^T L p = \frac 1 2 \sum_{(i, j) \in E} (p_i - p_j)^2,
\end{equation}
a small value of $p^TLp$ indicates that the absolute difference $|p_i-p_j|$ is small for all edges $(i,j)\in E$. This aligns with our assumption that the discovery probability remains fairly uniform across adjacent regions.

On the other hand, the true count of incidents $n$ should be determined by some region-related factors. We posit that the true counts follow a log-linear model \cite{von2012log}, a common choice for count data. This leads us to the following equation:
\begin{equation}
\log n = X\beta + \epsilon,
\end{equation}
where $n = (n_1, \cdots, n_M)^T\in\mathbb{N}^{M\times 1}$ is the vector of true counts, $\epsilon\sim N(0,\sigma_{n}^{2}I_M)$ is the error term, $\beta\in\mathbb{R}^{K\times 1}$ is the vector of parameters, and $X\in\mathbb{R}^{M\times K}$ is a known matrix representing the influence of the features on the incident counts. Typically we can think of $X$ as some random matrix. However, we need to notice that $X\beta$ is required to be greater than $0$ to  $n\ge 1$.

These mathematical formulations of our assumptions lay the groundwork for our subsequent analysis. They enable us to capture the nuances of the problem and develop a more rigorous and accurate model to estimate the true counts and discovery probabilities. Meanwhile, they lead to the final formulation of our problem.

Equipped with the two assumptions, we aim to develop an algorithm to address the under-count data issue. In the subsequent section, we will detail the nature of this problem and describe our solution.

The algorithm we propose employs the pre-established assumptions, incorporating the geographical data structure, influences of neighboring areas, and the potential socio-economic elements affecting the count of incidents. We will elaborate on the details of the problem, addressing its inherent difficulties and our proposed algorithm. This approach sets our work apart from earlier studies in this field, which tended to approach the issue from a more statistical perspective.

\section{Proposed debiasing algorithm: GRAUD}
\label{alg}
This section introduces an optimization problem as a part of a novel debiasing algorithm GRAUD. In our problem, both $n$ and $y$ represent count data, which suggests that a log transformation might be beneficial as it can make the data more normally distributed and reduce the variability \cite{changyong2014log}. Applying a log transformation to count data is a common practice in statistical analysis due to these advantages. We define the log transform of $n$, $p$ and $y$ as $u=\log n\in\R^M$, $v=\log p\in\R^M$ and $\tilde{y}=\log y\in\R^M$.  The formulation of optimization originates from the fact that $\mathbb{E}[y]=np$, which can be rewritten as $\tilde{y}\approx u+v$. This suggests minimizing the distance between $\tilde{y}$ and $u+v$. It is also worth noticing that by doing log transformation, we are able to minimize $\|\tilde{y}-u-v\|^2$ instead of $\|y-n\odot p\|^2$, where $\odot$ is elementwise (Hadamard) product. This turns the problem into a convex one.

As introduced in Section \ref{setting}, the graph Laplacian quadratic form $p^TLp$ is an ideal measure for graph smoothness. With log transformation, a similar term $v^TLV$ works in the same way. This is because
\begin{equation}
v^TLv = \frac{1}{2}\sum_{(i, j) \in E} (v_i - v_j)^2,
\end{equation}
which measures the differences between $v_i=\log p_i$. A small $v^TLv$ indicates closer $\log p_i$, which requires $p_i$ to be close to each other.

Combining the assumption that the true count $n$ follows a log-linear model, our proposed optimization problem is given by
\begin{equation}
\label{orgp}
\begin{aligned}
\min_{u,v,\beta} \|\tilde{y} - u -v\|^2 &+ \lambda_1 v^T Lv + \lambda_2 \|u-X\beta\|^2,\\
\text{s.t.}&\quad u\ge\tilde{y},v\le 0,    
\end{aligned}
\end{equation}
where $X$ is a known matrix, $\lambda_1$ and $\lambda_2$ are regularization constants. The constraint $u\ge\tilde{y}$ is because the true count $n$ should be larger than the observed count $y$, and $v\le 0$ is because $0\le p\le 1$.

In Equation \ref{orgp}, the optimal $\beta$ can be directly computed through least-squares regression:
\begin{equation}
\beta^* =\argmin_{\beta} \|n-X\beta\|^2
=(X^TX)^{-1}X^Tn.
\end{equation}

\begin{remark}
It is worth noticing that if the column space of $X$ is the full space, the term $\min_\beta \|n-X\beta\|^2$ will be zero, resulting in an identifiability issue as the log-linear model does not provide any additional information about $n$. Hence, it is necessary for $X$ not to be full row rank. Remember that each row of $X$ represents the influence of different factors on the true count $n$ for different regions, we can expect different rows to be similar or identical, which helps avoid $X$ to have full row rank. 
\end{remark}

Upon substituting the optimal $\beta^*$, the optimization problem transforms into:

\begin{equation}
\label{newp}
\begin{aligned}
\min_{u,v,\beta} \|\tilde{y} - u -v\|^2 &+ \lambda_1 v^T Lv + \lambda_2 u^THu,\\
\text{s.t.}&\quad u\ge\tilde{y},v\le 0,    
\end{aligned}
\end{equation}
where $H=I-X(X^TX)^{-1}X^T\in\R^{M\times M}$ is the projection matrix.

The gradients of $u$ and $v$ can be computed as follows:
\begin{equation}
2(u+v-\tilde{y}+\lambda_2 Hu),
\end{equation}
and
\begin{equation}
2(u+v-\tilde{y}+\lambda_1 Lv),
\end{equation}
respectively. Given these gradients, an alternating minimization algorithm can be utilized to address this optimization problem, and this proposed method is outlined in Algorithm \ref{alg1}.

If we take a close look at formulation \eqref{newp}, we can see that it is actually a graph signal separation problem in the sense that we try to separate the two signals $u$ and $v$ on a graph given observation $\tilde{y}$. Interestingly, this graph separation task aligns with a recent study \cite{10064133}, which also exploits graph smoothness in the process of signal demixing. Despite the similarity, it is critical to acknowledge the different points of departure of these two investigations. While both studies make use of graph smoothness, the previous work's theoretical result only emphasizes the aspect of convexity and uniqueness of the solution. On the contrary, our study goes beyond that and presents a more extensive theoretical analysis, containing recovery bound and convergence analysis.    

In the following section, we want to provide more insights and theoretical guarantees on the optimization problem we formulate and GRAUD.

\begin{algorithm}[t]
\caption{Graph-based debiasing Algorithm for Underreported Data (GRAUD)}
\label{alg1}
\begin{algorithmic}
\Require Initial $u_1$, $v_1$, $\tilde{y}$, $\lambda_1$, $\lambda_2$, $L$, $H$, iteration $T_{\text{in}}$, $T_{\text{out}}$, threshold $\epsilon$, stepsize $\eta$
\Ensure $n^* = \exp (u_{T_{\text{out}}})$, $p^* = \exp (v_{T_{\text{out}}})$
\For{$k = 1, \cdots, T_{\text{out}}$}:
    \For{$t = 1, \cdots, T_{\text{in}}$}:
        \State $du= u_k+v_k-\tilde{y}+\lambda_2 Hu_k$
        \State $u_{k} = u_{k} - \eta du$
    \EndFor
    \For{$t = 1, \cdots, T_{\text{in}}$}:
        \State $dv= u_{k}+v_k-\tilde{y}+\lambda_2 Lv_k$
        \State $v_{k} = v_{k} - \eta dv$
\EndFor
\State $u_{k+1}=u_{k}$, $v_{k+1}=v_{k}$
\EndFor
\end{algorithmic}
\end{algorithm}

\section{Theoretical Analysis}
\label{thy}
\subsection{Property of Binomial Distribution}
First, let's enumerate the assumptions vital to our approach. These assumptions direct the algorithm design and lead to theoretical guarantees. Here we denote $u_0=\log n_0$ and $v_0=\log p_0$ as ground truth.
\begin{assumption}
\label{a1}
Let $\epsilon_u = u_{0}^{T}Hu_0$ and $\epsilon_v = v_0^{T}Lv_0$, we assume that the quantity $\epsilon_u$ and $\epsilon_v$ are small.
\end{assumption}
\begin{assumption}
\label{a2}
$\nullop(L)\cap\nullop(H)=\{0\}.$
\end{assumption}

\begin{remark}
The terms in the first assumption measure the graph smoothness and the magnitude of noise in the log-linear model, which play a crucial role in the accuracy of GRAUD. The second assumption states that the only common element between the null spaces of $H$ and $L$ is the zero vector. This is important for resolving the identifiability issue because otherwise, an infinite number of solutions could be generated by merely adding and subtracting a common element to $u$ and $v$. Besides, we know $\rank(\nullop(L))=M-\rank(L)=1$ for a connected graph, and $\nullop(L)=\{c\cdot \bm{1}, c\in\R\}$. Assumption \ref{a2} satisfies as long as $\bm{1}\notin \nullop(H)$.
\end{remark}

The convex demixing theory \cite{mccoy2014sharp} is established upon similar ideas. This study demonstrates the exact recovery of signals under specific conditions related to the feasible cone. These conditions are generally met for sparse vectors or low-rank matrices. However, in our scenario, it is unlikely that either $u$ or $v$ has sparsity. This characteristic distinguishes our findings from those established results within the domain of convex demixing theory.

Before we delve into the theoretical assurances provided by GRAUD, it's necessary for us to explore some key results concerning the binomial distribution. In particular, we will draw upon a result derived from the Chernoff bound \cite{chernoff1952measure}. The Chernoff bound is an influential statistical concept used to derive a stringent upper limit on the probability that the aggregate of random variables strays from its anticipated value. It holds a notable place in probability theory and statistics, and is especially pertinent when handling large sums of independent random variables.

This theorem will be instrumental in helping us understand the distribution of the binomial data we're dealing with, providing a mathematical foundation for our work. The general form of Chernoff bound is as follows:
\begin{lemma}
\label{cheb}
(Chernoff bound) Let $y \sim \operatorname{Binomial}(n, p)$ and let $\mu=\mathbb{E}[y]$. For any $0<\delta<1$, we have the upper tail bound:
$$
\mathbb{P}(y \geq(1+\delta) \mu) \leq \exp \left(-\frac{\delta^2 \mu}{3}\right),
$$
and the lower tail bound:
$$
\mathbb{P}(y \leq(1-\delta) \mu) \leq \exp \left(-\frac{\delta^2 \mu}{2}\right).
$$
\end{lemma}

As a result, we have the following corollary. The proof can be found in the Appendix \ref{corp}.
\begin{lemma}
\label{c1}
If $y\sim\operatorname{Binomial}(n,p)$, then \[\mathbb{P}\left(|y-n p|\ge n^{1 / 2+\epsilon}\right)\le 2 e^{-n^{2 \epsilon}/3p}\] for small enough $\epsilon$.
\end{lemma}

If we do a log transformation to a binomial random variable $X$, then
\begin{equation}
\begin{aligned}
y-np&\ge n^{1/2+\epsilon} \\
\Leftrightarrow \log y \ge \log n + &\log p + \log(1 + n^{-1/2+\epsilon}/p) \\
y-np&\le n^{1/2+\epsilon} \\
\Leftrightarrow \log y \le \log n + &\log p - \log(1 - n^{-1/2+\epsilon}/p) 
\end{aligned}
\end{equation}
Thus a log-transformed version of the bound can be obtained as follows:
\begin{corollary}
If $y\sim\operatorname{Binomial}(n,p)$, then $\mathbb{P}\left(|\log y-\log n - \log p|\ge n^{-1 / 2+\epsilon/p}\right)\le 2 e^{-n^{2 \epsilon}/3p}$ for small enough $\epsilon$.
\end{corollary}
When the true incident count $n$ is large, the term $n^{-1 / 2+\epsilon/p}$ and $e^{-n^{2 \epsilon}/3p}$ all go to zero. The corollary shows that the random variable $\log y$ will highly concentrate around $\log n + \log p$. With the property introduced, we are able to derive recovery guarantee for our proposed optimization problem.

\subsection{Recovery Guarantee}
In this section, we present theoretical results concerning the uniqueness and the accuracy of the solution to our proposed optimization problem \ref{newp}.

Recall that $u = \log n\ge \log y=\tilde{y}$, $v = \log p \le 0$ and $\tilde{y}=\log y$, where $y\sim \operatorname{Binomial}(n,p)$. Proof of the results can be found in the Appendix.

\begin{proposition}
\label{cov}
Under assumption \ref{a2}, problem \eqref{newp} is convex and has a unique solution.
\end{proposition}

Let $\epsilon_y = \tilde{y} - u_0 - v_0$, and $\epsilon_u$, $\epsilon_v$ defined in Assumption \ref{a1}. The following is our main theorem.

\begin{theorem}
\label{ns}
Under assumption \ref{a1} and \ref{a2}, the solution $u^*$ and $v^*$ of the optimization problem \eqref{newp} satisfies 
\begin{equation}
\begin{aligned}
\|u^*-u_0\|^2\le \frac{2}{\delta_1}(2+\frac{1}{\sqrt{\lambda_1\lambda_{\min}(L)}}+\frac{1}{\sqrt{\lambda_2\lambda_{\min}(H)}})^2\|\epsilon_y\|^2+\frac{\epsilon_u}{\lambda_{\min}(H)}\\
\|v^*-v_0\|^2\le \frac{2}{\delta_1}(2+\frac{1}{\sqrt{\lambda_1\lambda_{\min}(L)}}+\frac{1}{\sqrt{\lambda_2\lambda_{\min}(H)}})^2\|\epsilon_y\|^2+\frac{\epsilon_v}{\lambda_{\min}(L)},
\end{aligned}
\end{equation}
where $\delta_1$ is the smallest singular value of the block matrix $\begin{bmatrix}
L & H
\end{bmatrix}\in\R^{M\times 2M}$, and $\lambda_{\min}(X)$ denotes the smallest non-zero eigenvalue of a square matrix $X$.
\end{theorem}

We have the following corollary demonstrating exact recovery under the "noiseless" condition.
\begin{corollary}
\label{nsl}
Under assumption \ref{a1} and \ref{a2}, and $\epsilon_y=\epsilon_u=\epsilon_v=0$, the optimization problem \eqref{newp} has exact recovery $u^* = u_0$, $v^* = v_0$, where $u^*$ and $v^*$ are solutions of the problem.
\end{corollary}

The theorem, denoted as \eqref{ns}, offers a recovery limit concerning the optimization problem expressed in \eqref{newp}. This recovery bound is a key measure as it reflects the potential effectiveness and accuracy of our proposed solution. It serves as a performance metric for how closely the solution we obtain aligns with the true optimal solution.

Our earlier analysis indicates that the term $\|\epsilon_y\|$ diminishes towards zero with a high probability as $n$ increases. This is an encouraging trend because it shows that the discrepancy in our solution decreases as the count grows. Additionally, based on our initial assumptions, the terms $\|\epsilon_1\|$ and $\|\epsilon_2\|$ are expected to be very small. Given these factors, we can infer that the upper bound delineated on the right side of the equation will be considerably small. This signifies a high quality of the solution, indicating that our optimization strategy can solve the problem effectively under the given assumptions.

We can see that our optimization problem is block multiconvex for $u$ and $v$. Besides, the objective function is strongly convex for $u$ and for $v$. Considering that the objective is a real analytic function in $\R^{2M}$, it satisfies the Kurdyka–Lojasiewicz (KL) inequality. From the results regarding the block coordinate descent method in \cite{xu2013block}, a global convergence result can be achieved for our Algorithm \ref{alg1}. It follows that the algorithm converges sublinearly.

\begin{proposition}
The output $(u_k, v_k)$ generated by Algorithm \ref{alg1} converges to a critical point of Problem \eqref{newp}, which is the unique global minimum.
\end{proposition}

\section{Numerical Experiments}
\label{exp}
\subsection{Simulated Examples}
We proceed to evaluate the efficacy of GRAUD through two simulated examples. In the initial experiment, we arbitrarily select $X=2+Z\in\mathbb{R}^{M\times K}$, where $Z$ is a standard normal matrix, and set $\beta\in\mathbb{R}^{K\times1}$ as an all-one vector. We adopt $M=10$ and $K=3$ for this experiment. Then, we create $n = [\exp(X\beta)]$, ensuring that Assumption \eqref{a1} is satisfied by maintaining $n^THn$ relatively small. For $p$, we generate it through equation $p = 0.7+0.1\epsilon$, where $\epsilon\sim N(0,I_M)$ is from a standard normal distribution. We subsequently constrain $p$ to fall within the $[0.05,0.95]$ interval to circumvent extreme scenarios. Furthermore, we set $p^TLp\le 0.02$ to satisfy Assumption \ref{a1}. $\lambda_1$ is selected to be 0.01, and $\lambda_2$ is 0.9. We select the regularization parameters through $5$-fold cross-validation. For the initial values, we assign $n_1 = y$ and $p_1 = y/n_1$, which is a natural choice. This results in $u_1= \log n_1$ and $v_1 = \log p_1$.

\begin{figure}[t]
\centering
\includegraphics[width=0.45\textwidth]{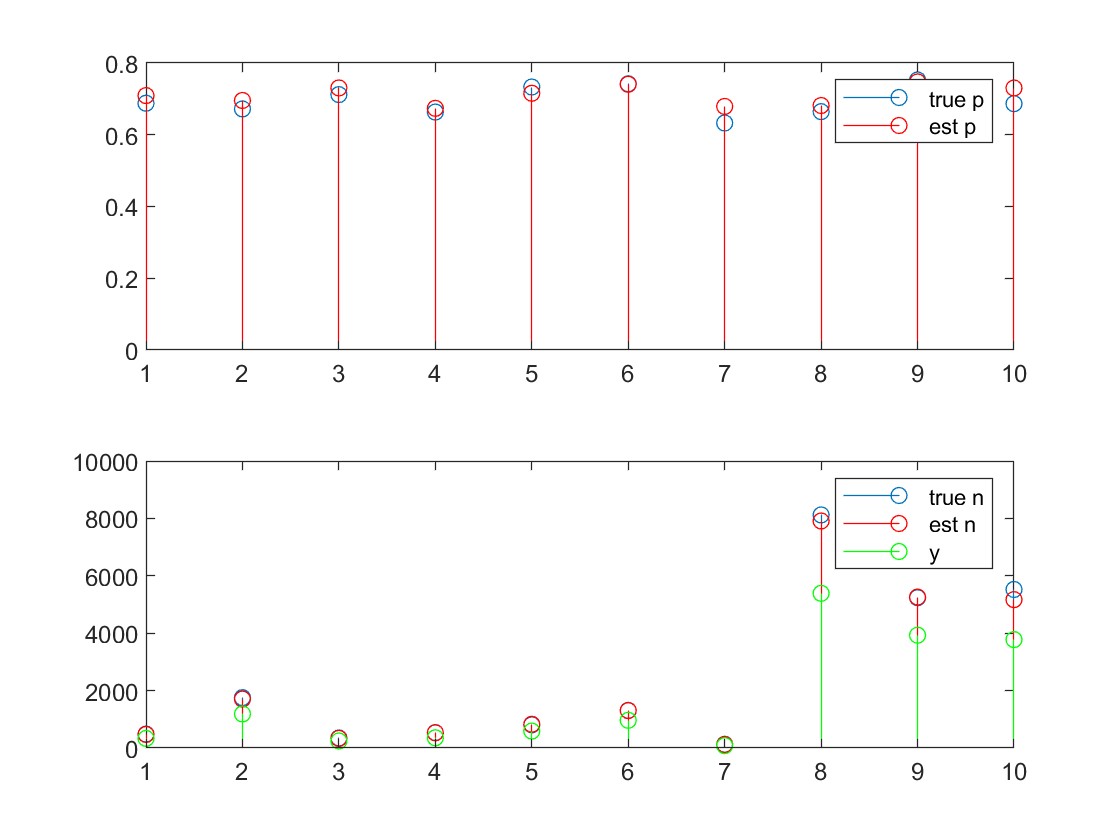}
\includegraphics[width=0.45\textwidth]{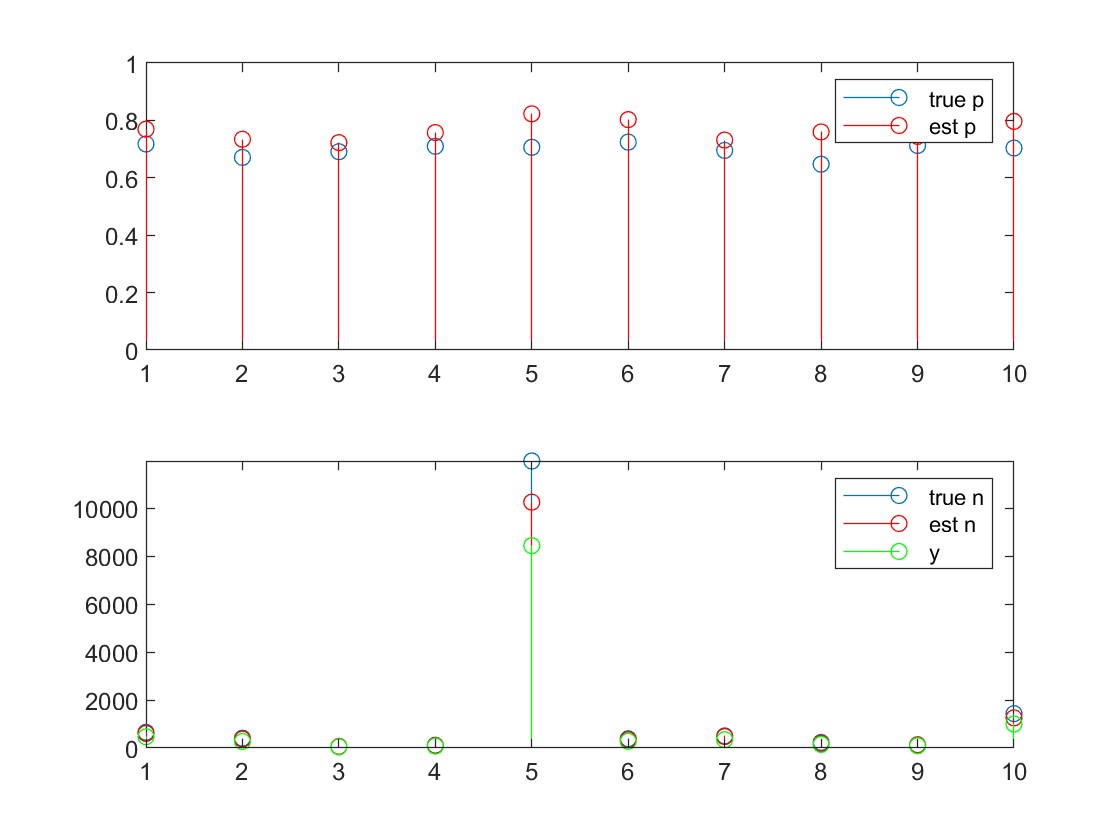}
\caption{The two plots on the left present a comparison between the true value of $p$ and its estimated counterpart, as well as a comparison between the true value of $n$, the estimated value of $n$, and $y$ for the first sample. The relative $\ell_1$ error for $p$ stands at $0.035$, while the relative $\ell_1$ error for $n$ is $0.029$. The two plots on the right reveal the same comparison for the second sample. In this case, the relative $\ell_1$ error for $p$ is $0.095$ and the relative $\ell_1$ error for $n$ is $0.132$.}
\label{fig:e1}
\end{figure}

\begin{figure}[t]
\centering
\includegraphics[width=0.5\textwidth]{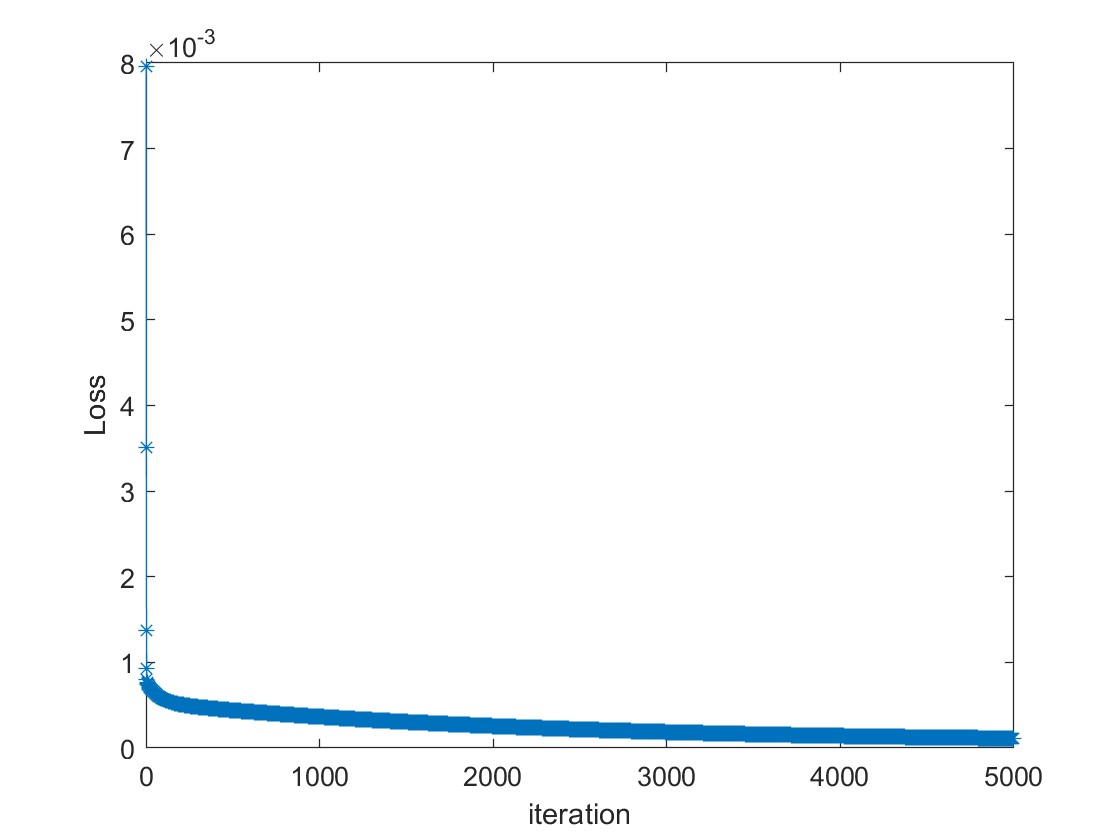}
\caption{The figure shows the rapid decrease of loss \eqref{newp} in GRAUD for simulated data.}
\label{fig:c1}
\end{figure}

As demonstrated in Figure \ref{fig:e1}, the debiased solution generated by GRAUD is closely aligned with the ground truth, indicating high accuracy in our approach. The proximity of GRAUD's output to the ground truth underscores its reliability in providing accurate results, thus justifying its application in this context. Furthermore, Figure \ref{fig:c1} exhibits the rapid convergence rate of GRAUD, reinforcing its practical utility. The rapid convergence, as depicted in this figure, does not rely on the initial guess of the algorithm.

\begin{figure}[t]
    \centering
    \includegraphics[width =.45\linewidth]{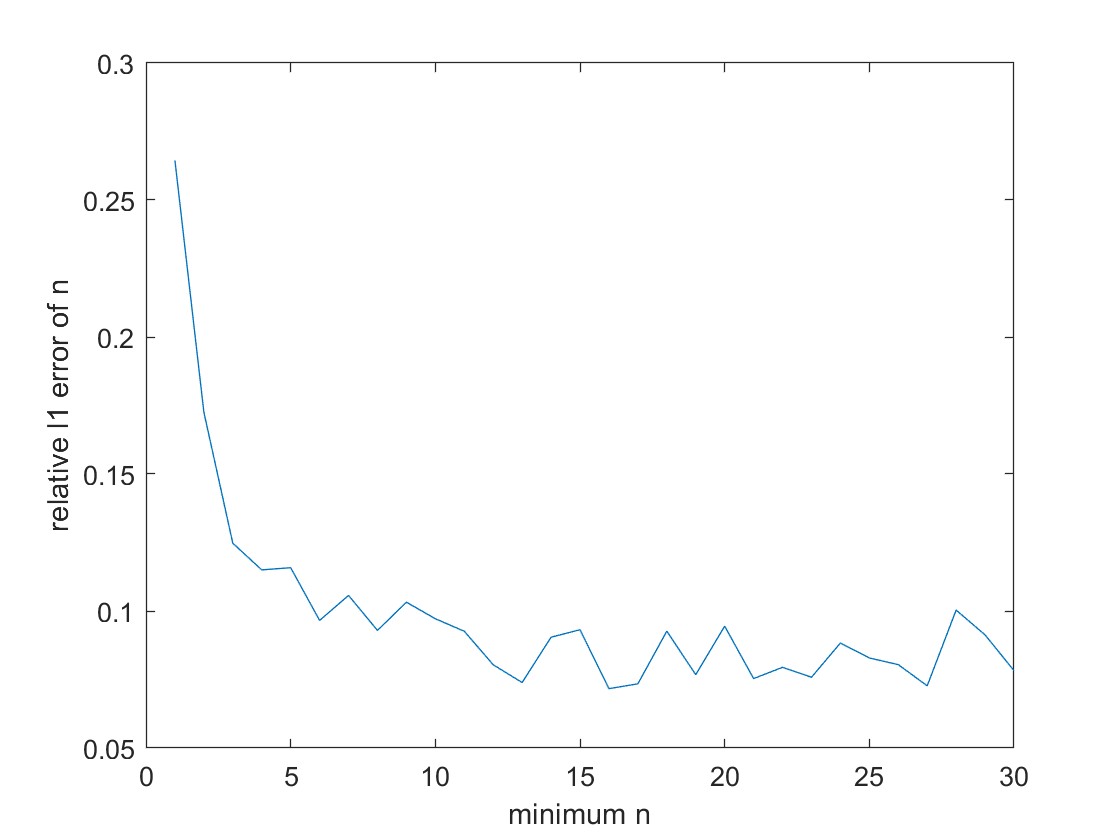}
    \includegraphics[width =.45\linewidth]{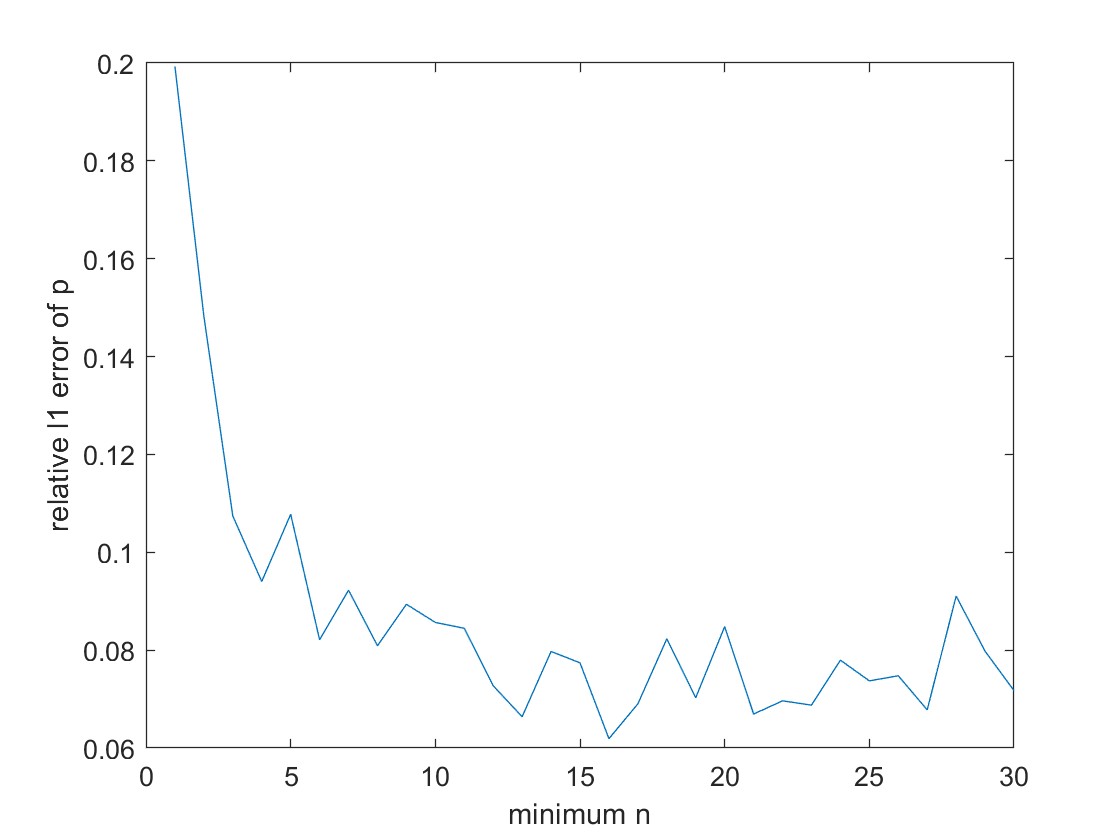}
    \caption{The figure shows the relative $\ell_1$ loss of $n$ and $p$ with the growth of $n$.}
    \label{fig:changen}
\end{figure}

We further examine the relationship between the algorithm's accuracy and the true count $n$. Specifically, we create a sample where the minimum $n_i$ is a given value $k$. We then apply GRAUD to this sample and calculate the relative $\ell_1$ error for $n$ and $p$. The procedure is repeated 100 times, and we derive an average error for each $k$ in the range from 1 to 30. The resulting loss pattern is depicted in Figure \ref{fig:changen}.

The chart demonstrates a decline in loss as $n$ increases, reaching a steady state once $n$ surpasses $10$. From this observation, it is evident that GRAUD can accurately estimate both $n$ and $p$. This trend conforms to our theoretical findings that indicate a larger $n$ yields less "noise", thus enhancing the precision of GRAUD.

Moreover, the experimental results suggest that an $n$ value exceeding 10 is large enough to achieve a dependable estimate. This threshold is readily achievable under real-world scenarios, emphasizing the feasibility of our approach.

\begin{figure}[t]
    \centering
    \includegraphics[width =.45\linewidth]{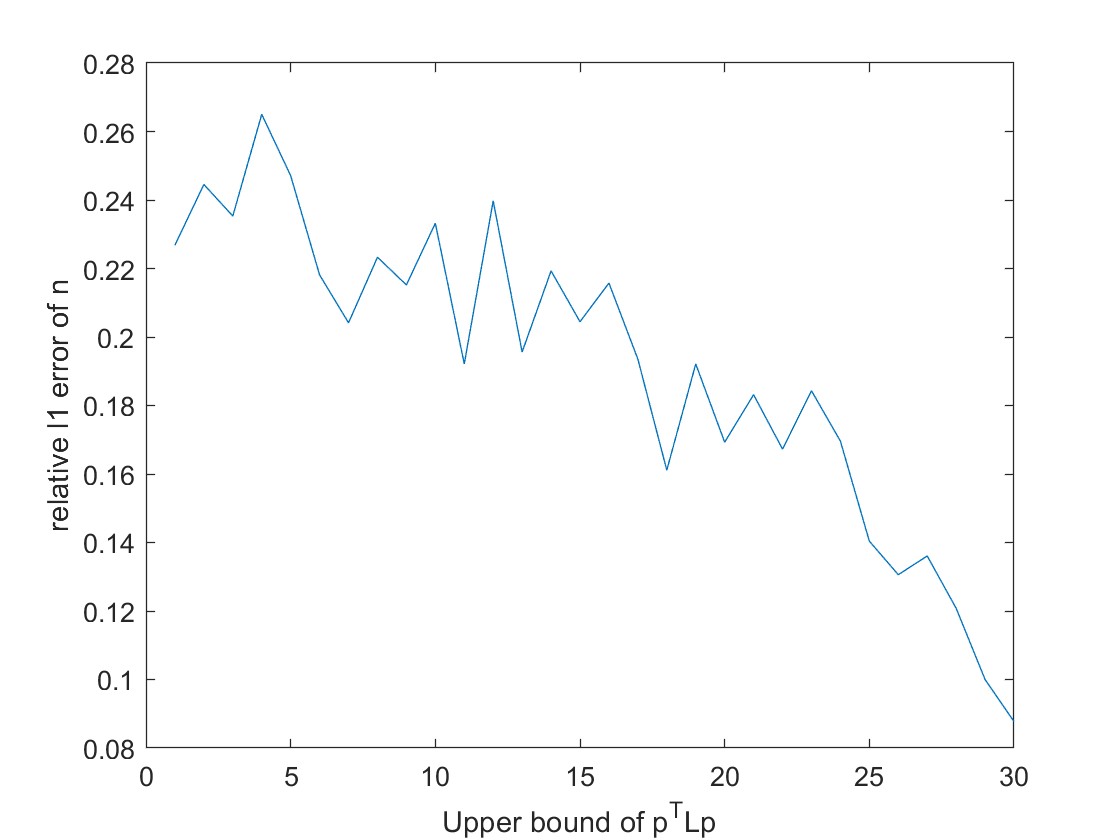}
    \includegraphics[width =.45\linewidth]{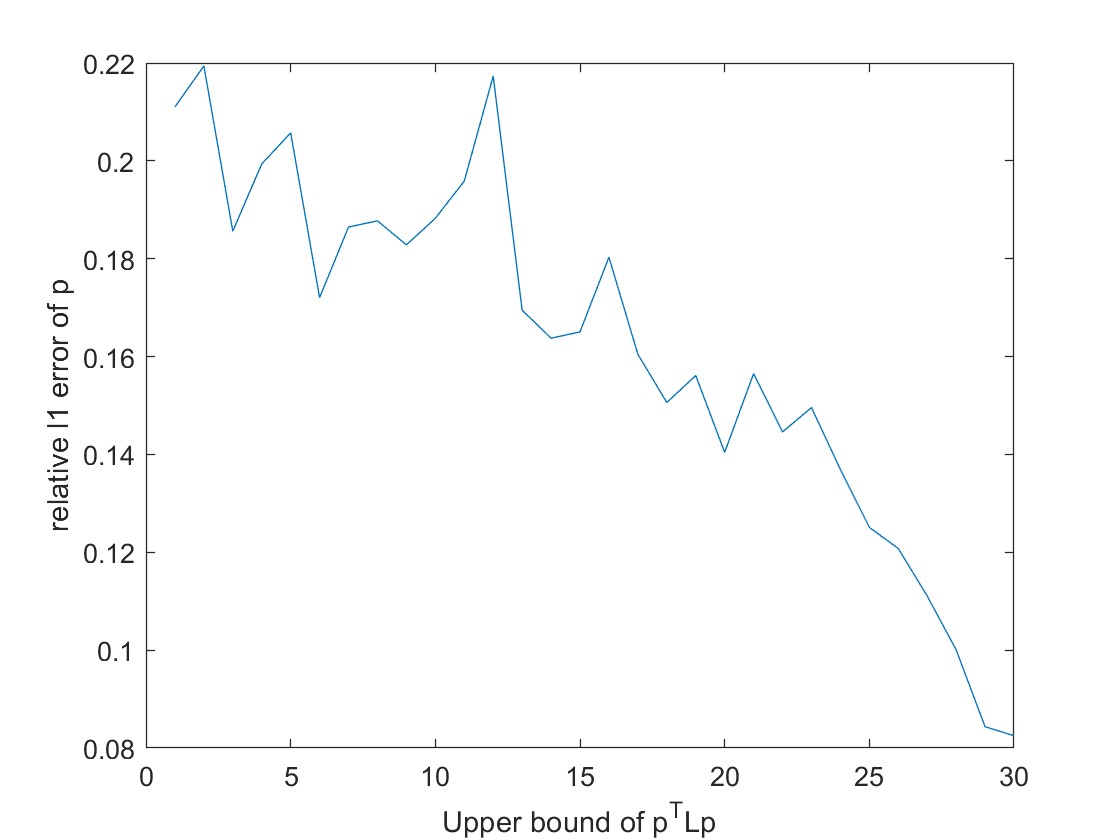}
    \caption{The figure shows the relative $\ell_1$ loss of $n$ and $p$ with the decrease of $p$.}
    \label{fig:changep}
\end{figure}

In a similar manner, we study the correlation between the accuracy of GRAUD and the true discovery probability $p$. We keep the value of $p^TLp/M$ below $0.001k$ for $k$ ranging from 1 to 30. Figure \ref{fig:changep} illustrates the loss trend as $p^TLp$ decreases. It can be seen that the loss steadily diminishes as $p^TLp$ becomes smaller, which is also in alignment with our theoretical expectations.

We also tried a harder case where $m=20$. The number of nodes is getting larger and thus harder to recover. We can still generate similar results. In this case we set $p = 0.3+0.05\epsilon$, where $\epsilon\sim N(0,I_M)$ is a standard normal noise. We stick to the criterion of keeping $p^TLp/M$ below $0.001$. Overall, the relative $\ell_1$ error of GRAUD is around $0.08$, which is similar to the accuracy in the previous experiment.

\begin{figure}[t]
\centering
\includegraphics[width=0.45\textwidth]{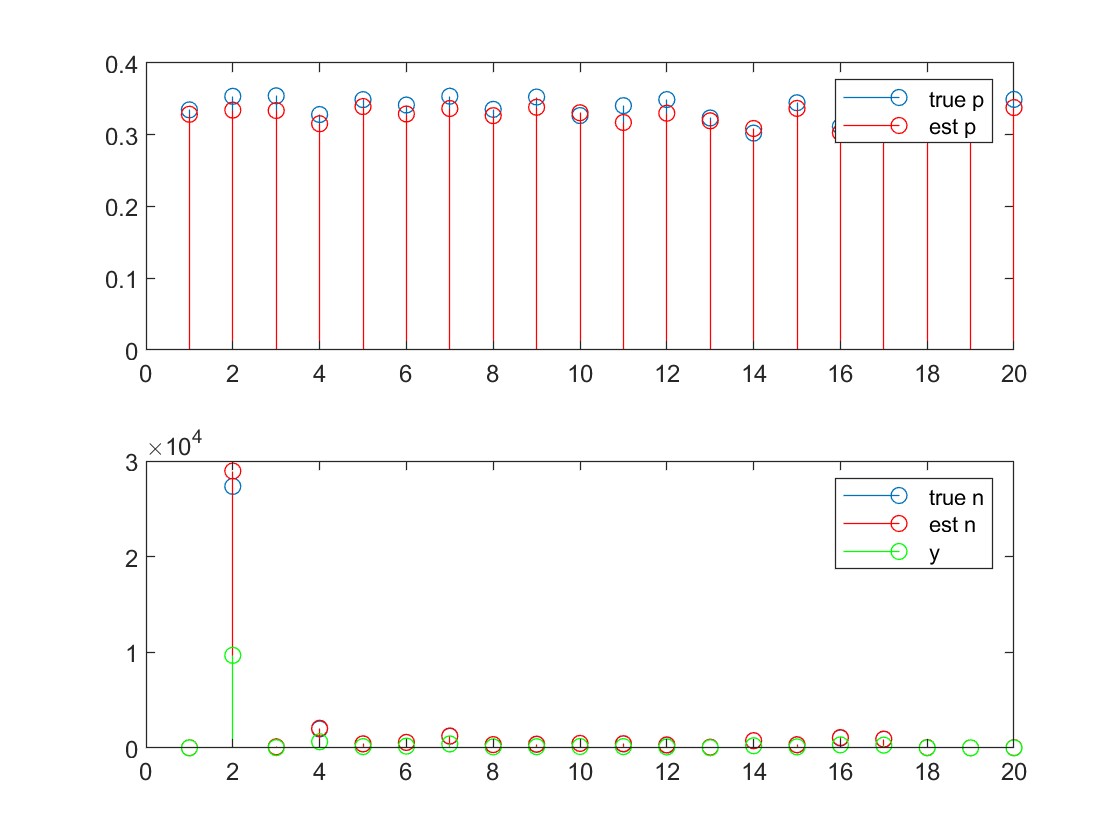}
\includegraphics[width=0.45\textwidth]{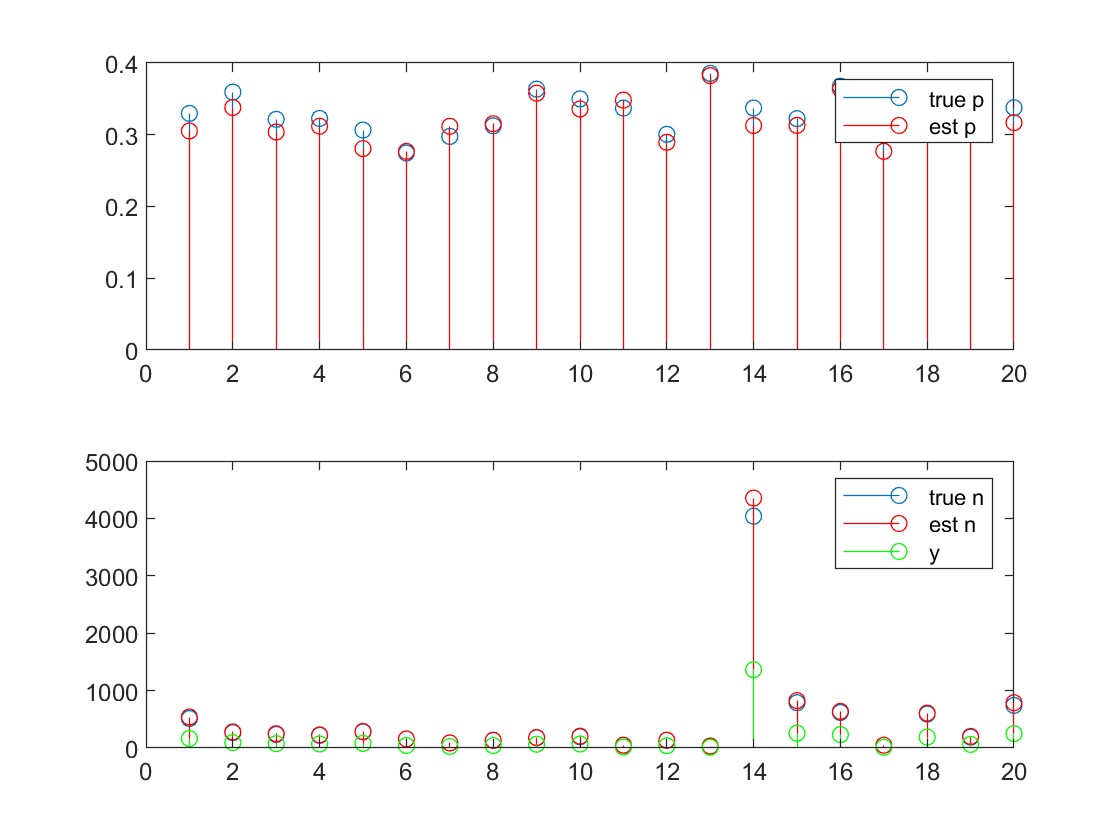}
\caption{The first two plots present a comparison between the true value of $p$ and its estimated counterpart, as well as a comparison between the true value of $n$, the estimated value of $n$, and $y$ for the first sample. The relative $\ell_1$ error for $p$ stands at $0.035$, while the relative $\ell_1$ error for $n$ is $0.052$. The final two plots reveal the same comparison for the second sample. In this case, the relative $\ell_1$ error for $p$ is $0.043$ and the relative $\ell_1$ error for $n$ is $0.056$.}
\label{fig:e2}
\end{figure}

\subsection{Real Data Experiment}
\begin{figure}[t]
    \centering
    \includegraphics[width =.45\linewidth]{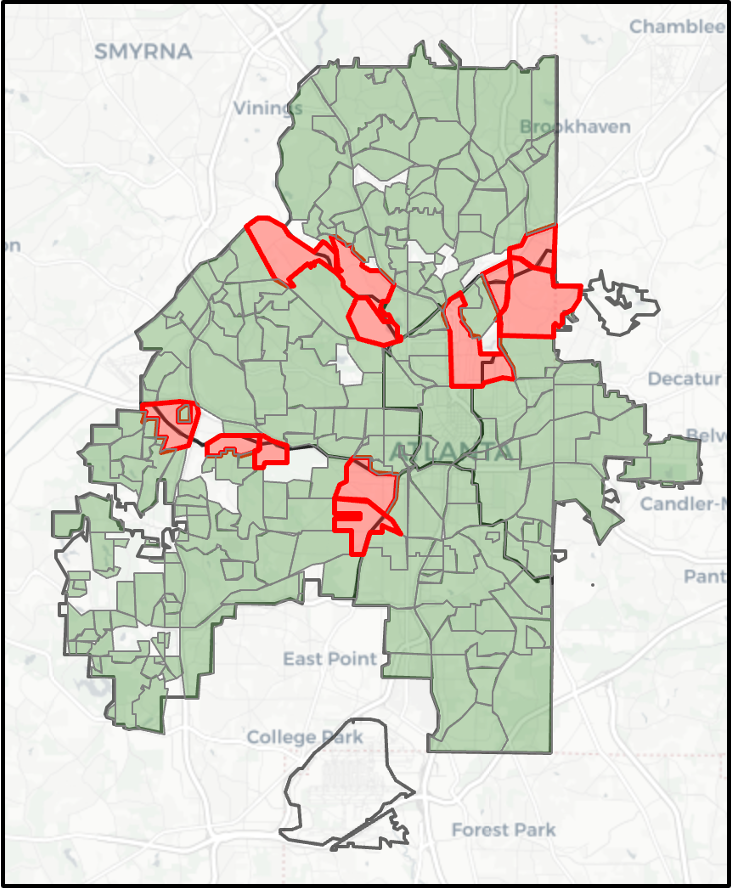}
     \includegraphics[width =.45\linewidth]{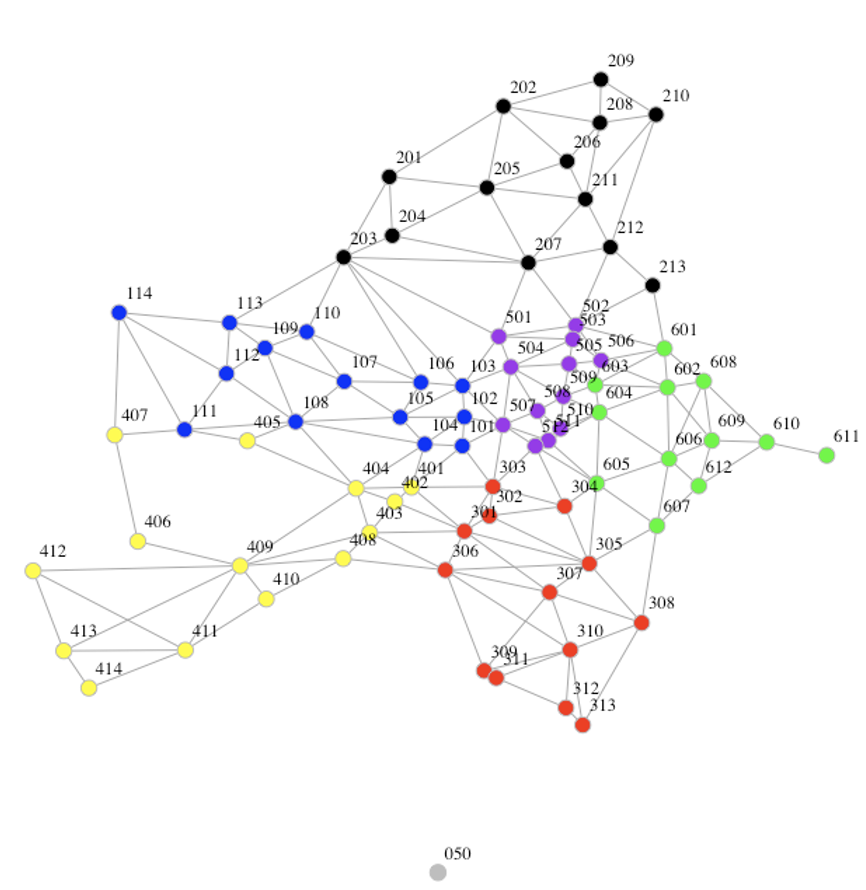}
    \caption{The first figure shows how Atlanta is divided into $78$ distinct beats. The second plot demonstrates the associated graph of beats. The number of each node is the corresponding code for the beat.}
    \label{fig:APD}
\end{figure}

In our real-world experiment, we direct our attention towards emergency (911) call data originating from Atlanta, specifically from the year 2019, which comprises approximately 580,000 instances. It's noteworthy that the actual number of emergency situations is likely higher than represented by these calls, as they tend to underestimate the true magnitude of emergencies. Our objective is to correct this under-reporting, enabling a more accurate estimate of the true emergency cases. The inherent graph relationship in this data and the approximate satisfaction of our previously mentioned assumptions make it a suitable subject for our analysis. We make use of this data to establish the variable $y_i$ for every individual beat, with a beat referring to the distinct geographical area assigned to a police officer for patrolling. These beats subdivide Atlanta into 78 distinct sections, as illustrated in Figure \ref{fig:APD}, which offers a naturally discrete geographical division for our research.

To enhance our understanding, we create a graphical model in which each beat is symbolized as a node, and edges are formed between nodes that correspond to neighboring beats. This graph-based representation allows us to visualize and comprehend the spatial connections and proximity among the different beats in a more intuitive manner.

To supplement our dataset further, we include the census data from 2019, factoring in aspects such as population size, income, and level of education (quantified as the fraction of the population that has achieved at least a high school diploma). These factors constitute our $x_{ij}$ variables, thereby incorporating socioeconomic factors into our analysis.

\begin{figure}[t]
    \centering
    \includegraphics[width =.45\linewidth]{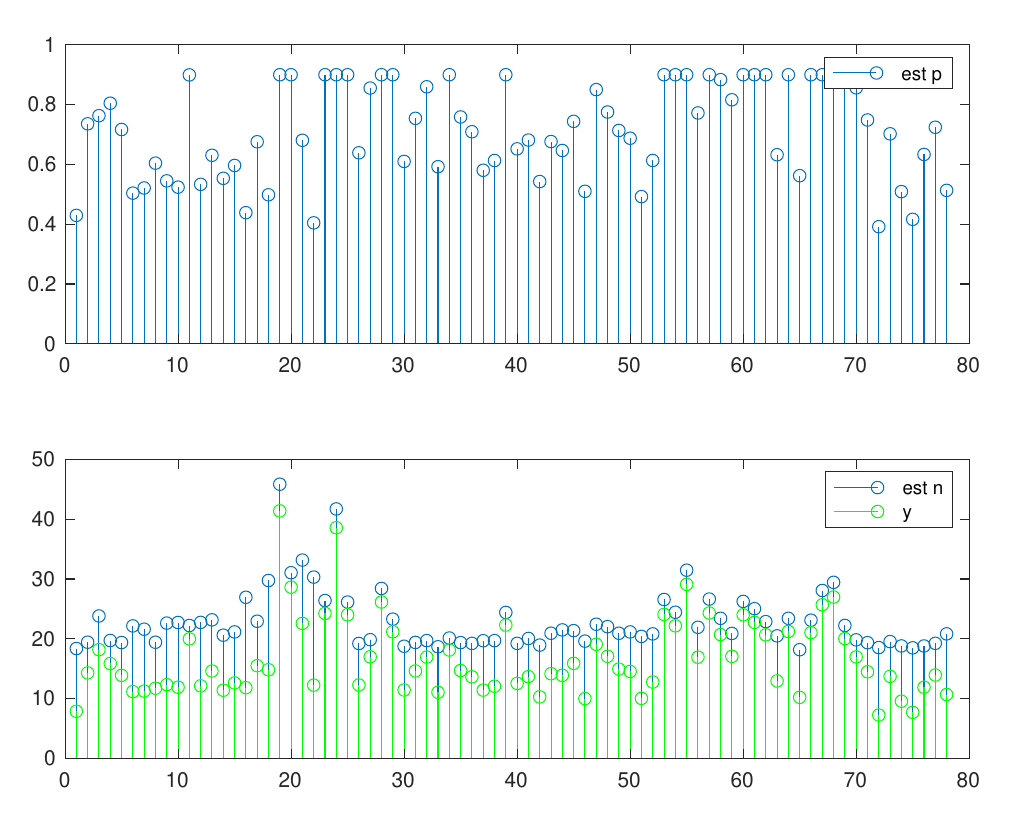}
    \caption{The estimated $\hat p_i$ from data (when converges to locally). The $p_i$ are intialized to be constant 0.1. Clearly, the different regions may have different level of crime discovery.}
    \label{fig:p_hat}
\end{figure}

To begin our analysis, we set the $p_i$ vector as a vector of all 0.8s and $n_i=y_i/p_i$. The localized solution we achieve from this starting point is represented in Figure \ref{fig:p_hat}. The yellow areas represent a higher discovery probability, and those areas with higher discovery rates are mainly located in the downtown, midtown or other prosperous areas in Atlanta. This makes sense because those flourishing areas usually have better public security and thus resulting in higher discovery rates.

\begin{figure}[t]
    \centering
    \includegraphics[width =.45\linewidth]{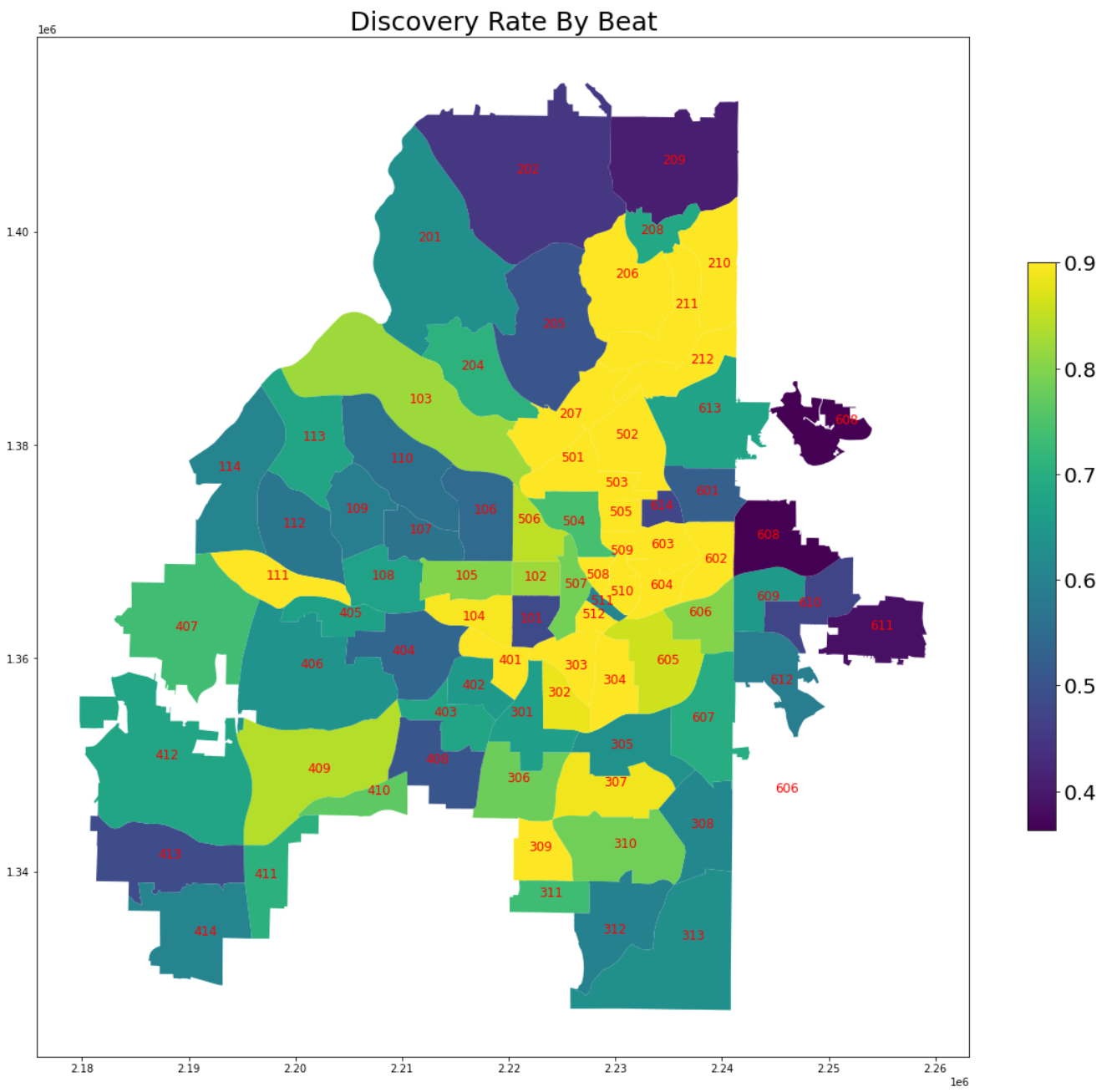}
    \caption{The estimated $p_i$ in each beat when initializing with the discovery probability of all 0.8.}
    \label{fig:my_label}
\end{figure}

\section{Conclusion}
\label{con}
In this paper, we proposed a novel graph prediction method for debiasing under-count data. The idea is to utilize the intrinsic graph structure of the problem and thus overcome the identifiability issue. We reformulate the problem as a constrained convex optimization problem and establish the connection between the binomial n problem and the graph signal separation problem. We provide an alternating minimization optimization algorithm for efficiently recovering the true count. Recovery bounds and convergence results are also established for our proposed method. We approach the binomial n problem from a novel perspective, contrasting with traditional methods, and view it through the lens of signal processing. Both the synthetic data experiment and the real data experiment demonstrate the accuracy and efficiency of our proposed method.

For future work, we would like to broaden our current framework to accommodate a more general context and refine our algorithm to enhance its resistance to noise. At present, the precision of GRAUD is contingent upon the fulfillment of certain assumptions, which might grow challenging as the dimensionality (or the number of nodes) increases. Therefore, we also anticipate relaxing these assumptions and improving the algorithm's functionality in higher dimensions.

\section*{Acknowledgements}
We would like to thank Sarah for her help with the real data experiment. This work is partially supported by an NSF CAREER CCF-1650913, and NSF DMS-2134037, CMMI-2015787, CMMI-2112533, DMS-1938106, and DMS-1830210, and a Coca-Cola Foundation fund.

\vspace{-0.1in}
\bibliography{reference}
\newpage
\section*{Appendix}

\subsection{Proof of Lemma \ref{c1}}
\label{corp}
From Lemma \ref{cheb}, we have
\begin{equation}
\begin{aligned}
\mathbb{P}(X\ge np+n^{1/2+\epsilon}) &=  \mathbb{P}(X\ge np(1+n^{-1/2+\epsilon}p^{-1}))\\
&\le \exp(- \frac{(n^{-1/2+\epsilon}p^{-1})^2}np{3})\\
&= \exp(-\frac{n^{2\epsilon}}{3p}).
\end{aligned}
\end{equation}
Similarly, we have the lower tail bound:
\begin{equation}
\mathbb{P}(X\le np-n^{1/2+\epsilon})\le\exp(-\frac{n^{2\epsilon}}{2p}).
\end{equation}
Combining the two inequalities, we have
\begin{equation}
\begin{aligned}
\mathbb{P}\left(|X-n p|\ge n^{1 / 2+\epsilon}\right) = &\mathbb{P}(X\ge np+n^{1/2+\epsilon}) +
\mathbb{P}(X\le np-n^{1/2+\epsilon})\\
\le &\exp(-\frac{n^{2\epsilon}}{3p})+\exp(-\frac{n^{2\epsilon}}{2p})\\
\le &2\exp(-\frac{n^{2\epsilon}}{3p}).
\end{aligned}
\end{equation}

\subsection{Proof of Proposition \ref{cov}}
First, we show that the objective function is convex. Let $L(u,v)$ be the loss function, we have
\begin{equation}
\frac{\partial L}{\partial u} = 2(u+v-\tilde{y}+\lambda_2 Hu),
\end{equation}
and
\begin{equation}
\frac{\partial L}{\partial v} = 2(u+v-\tilde{y}+\lambda_1 Lv).
\end{equation}
We can compute the Hessian matrix $\tilde{H}$:
\begin{equation}
\left[ 
\begin{array}{c c} 
  I+\lambda_2 H & I \\ 
  I & I+\lambda_1 L
\end{array} 
\right] 
\end{equation}
For any $x = (u,v)^T$, we have
\begin{equation}
\begin{aligned}
x^T\tilde{H}x &=u^T(I+\lambda_2 H)u +2u^T v + v^T(I+\lambda_1 L)v\\
&= \|u+v\|^2 + \lambda_2 u^THu + \lambda_1 v^TLV \ge 0.
\end{aligned}
\end{equation}
The inequality holds because $H$ and $L$ are positive semidefinite, which infers that $\tilde{H}$ is positive semidefinite, and leads to the convexity of the problem.

Then we will show the uniqueness of the solution. If this does not hold, then there will be two different solutions $x = (u, v)$ and $x^* = (u^*, v^*)$, and the gradient should be zero at those two points, which means
\begin{equation}
\begin{aligned}
\lambda_1 Lv &= \lambda_2 Hu = \tilde{y} - u -v\\
\lambda_1 Lv^* &= \lambda_2 Hu^* = \tilde{y} - u^* -v^*.
\end{aligned}
\end{equation}
Define $\delta_v^*=v-v^*$, $\delta_u^*=u-u^*$, then
\begin{equation}
\begin{aligned}
\lambda_1 L\delta_v &= \lambda_2 H\delta_u = 0 - \delta_u -\delta_v.
\end{aligned}
\end{equation}
We can rewrite the equation into a quadratic optimization problem:
\begin{equation}
\min_{\delta_v, \delta_u}\ \ f(\delta_v,\delta_u)=\|\delta_v+\delta_u\|^2+\lambda_1 \delta_v^TL\delta_v + \lambda_2 \delta_u^TH\delta_u.
\end{equation}
$\delta_v^*$ and $\delta_u^*$ is a non-zero solution of this problem. We notice that $0$ is a solution of the problem and the minimum should be $0$. This means $ f(\delta_v^*,\delta_u^*) = 0$. Considering the non-negativity of each term, we have $\|\delta_v^*+\delta_u^*\|=0$, $\delta_v^{*T}L\delta_v^* = 0$ and $\delta_u^{*T}H\delta_u^*=0$. This means $\delta_v^*$ is in the null space of $L$ and $\delta_u^*$ is in the null space of $H$. According to Assumption \eqref{a2}, the intersection of two null spaces is $0$, which means $\delta_v^*=\delta_u^*=0$. So the solution is unique.

\subsection{Proof of Theorem \ref{ns}}

Suppose $\tilde{y} = u_0 + v_0 + \epsilon_y$, $v_0 = h + \epsilon_1$, and $u_0 = X\beta_0 + \epsilon_2$. Here the subscript $0$ means the ground truth, and $h$ is the projection of $v_0$ in the null space of $L$.

From the assumption \eqref{a1}, we know that $\epsilon_v=\epsilon_1^TL\epsilon_1$ and $\epsilon_u=\epsilon_2^TH\epsilon_2$ are small and we can treat them as a kind of ``noise''. Denote $u-u_0 = d_u$, $v-v_0=d_v$, then we can rewrite the original objective function \eqref{newp}:
\begin{equation}
\begin{aligned}
&\|\tilde{y} - u - v\|^2 + \lambda_1 v^T Lv + \lambda_2 u^THu\\
=& \|u_0 + v_0 + \epsilon_y - (d_u + u_0) - (d_v + v_0)\|^2 +\lambda_1 (d_v+h+\epsilon_1)^TL(d_v+h+\epsilon_1) +\\
&\lambda_2 (d_u+X\beta_0+\epsilon_2)^TH(d_u+X\beta_0+\epsilon_2)\\
=&\|\epsilon_y -d_u-d_v\|^2 + \lambda_1 (d_v+\epsilon_1)^TL(d_v+\epsilon_1)+ \lambda_2 (d_u+\epsilon_2)^TH(d_u+\epsilon_2).
\end{aligned}
\end{equation}
If we define $\tilde{d}_v = d_v + \epsilon_1$, $\tilde{d}_u = d_u + \epsilon_2$ and $\epsilon = \epsilon_y + \epsilon_1 + \epsilon_2$, then the previous optimization problem is equivalent to
\begin{equation}
\min_{\tilde{d}_v,\tilde{d}_u} L(\tilde{d}_v,\tilde{d}_u)=\|\epsilon-\tilde{d}_v-\tilde{d}_u\|^2 + \lambda_1 \tilde{d}_v^TL\tilde{d}_v + \lambda_2 \tilde{d}_u^TH\tilde{d}_u.
\end{equation}
Let $\tilde{d}_v = \tilde{d}_u = 0$, the objective function $L(\tilde{d}_v,\tilde{d}_u)=\|\epsilon\|^2$. This means that $L(\tilde{d}_v^*,\tilde{d}_u^*)\le \|\epsilon\|^2$. Consequently, we have
\begin{equation}
\begin{aligned}
\label{i1}
\tilde{d}_v^{*T}L\tilde{d}_v^* \le \|\epsilon\|^2/\lambda_1\\
\tilde{d}_u^{*T}H\tilde{d}_u^* \le \|\epsilon\|^2/\lambda_2.
\end{aligned}
\end{equation}
Define $\lambda_{\min}(L)$ as the smallest positive eigenvalue of matrix $L$, and $\lambda_{\min}(H)$ as the smallest positive eigenvalue of matrix $H$. We can divide $\tilde{d}_v= \tilde{d}_{v\parallel}+\tilde{d}_{v\perp}$ and $\tilde{d}_u= \tilde{d}_{u\parallel}+\tilde{d}_{u\perp}$. Here $\tilde{d}_{v\parallel}$ is in the null space of $L$, $\tilde{d}_{v\perp}$ is orthogonal to the null space of $L$, $\tilde{d}_{u\parallel}$ is in the null space of $H$ and $\tilde{d}_{u\perp}$ is orthogonal to the null space of $H$,   We have
\begin{equation}
\label{i2}
\begin{aligned}
\tilde{d}_v^TL\tilde{d}_v \ge \lambda_{\min}(L) \|d_{v\perp}\|^2\\
\tilde{d}_u^TH\tilde{d}_u \ge \lambda_{\min}(H) \|d_{u\perp}\|^2.
\end{aligned}
\end{equation}
Combining \eqref{i1} and \eqref{i2}, we have $\|\tilde{d}_{v\perp}\|^2\le\frac{\|\epsilon\|^2}{\lambda_1\lambda_{\min}(L)}$ and $\|\tilde{d}_{u\perp}\|^2\le\frac{\|\epsilon\|^2}{\lambda_2\lambda_{\min}(H)}$. Next we would like to bound $\|\tilde{d}_{v\parallel}\|$ and $\|\tilde{d}_{u\parallel}\|$. Since $\tilde{d}_{v\parallel}$ is in the column space of $X$, there exists $a_1, a_2, \cdots, a_{r_X}$ s.t. $\tilde{d}_{v\parallel} = \sum_{i=1}^{r_X}a_i x_i$, where $x_1, \cdots, x_{r_X}$ be the orthonormal basis of the column space of $X$. Similarly, there exists $b_1, b_2, \cdots, b_{r_L}$ s.t. $\tilde{d}_{u\parallel} = \sum_{i=1}^{r_L}b_i l_i$, where $l_1, \cdots, l_{r_L}$ be the orthonormal basis of the null space of $L$. It is obvious that $\|\tilde{d}_{v\parallel}\|^2=\sum_{i=1}^{r_X}a_i^2$ and $\|\tilde{d}_{u\parallel}\|^2=\sum_{i=1}^{r_L}b_i^2$. From Assumption \eqref{a2}, we have
\begin{equation}
\begin{aligned}
\|\tilde{d}_{v\parallel}+\tilde{d}_{u\parallel}\|^2&\ge \delta_1 (\sum_{i=1}^{r_X}a_i^2+\sum_{i=1}^{r_L}b_i^2)\\
&=\delta_1(\|\tilde{d}_{v\parallel}\|^2 + \|\tilde{d}_{u\parallel}\|^2).
\end{aligned}
\end{equation}
Since $L(\tilde{d}_v^*,\tilde{d}_u^*)\le \|\epsilon\|^2$, we also have $\|\epsilon-\tilde{d}_v-\tilde{d}_u\|\le\|\epsilon\|$. Meanwhile,
\begin{equation}
\begin{aligned}
\|\epsilon\|&\ge\|\epsilon-\tilde{d}_v-\tilde{d}_u\|=\|\epsilon-(\tilde{d}_{v\parallel}+\tilde{d}_{v\perp})-(\tilde{d}_{u\parallel}+\tilde{d}_{u\perp})\|\\
&\ge\|\tilde{d}_{v\parallel}+\tilde{d}_{u\parallel}\|-\|\epsilon\|-\|\tilde{d}_{v\perp}\|-\|\tilde{d}_{u\perp}\|\\
&\ge \delta_1^{1/2}(\|\tilde{d}_{v\parallel}\|^2 + \|\tilde{d}_{u\parallel}\|^2)^{1/2}-\|\epsilon\|-\frac{\|\epsilon\|}{\sqrt{\lambda_1\lambda_{\min}(L)}}-\frac{\|\epsilon\|}{\sqrt{\lambda_2\lambda_{\min}(H)}}.
\end{aligned}
\end{equation}
This means
\begin{equation}
\delta_1^{1/2}(\|\tilde{d}_{v\parallel}\|^2 + \|\tilde{d}_{u\parallel}\|^2)^{1/2}\le \|\epsilon\|(2+\frac{1}{\sqrt{\lambda_1\lambda_{\min}(L)}}+\frac{1}{\sqrt{\lambda_2\lambda_{\min}(H)}}).
\end{equation}
Define $c_0 = 2+\frac{1}{\sqrt{\lambda_1\lambda_{\min}(L)}}+\frac{1}{\sqrt{\lambda_2\lambda_{\min}(H)}}$, we have
\begin{equation}
\|\tilde{d}_{v\parallel}\|^2 + \|\tilde{d}_{u\parallel}\|^2\le \frac{c_0^2\|\epsilon\|^2}{\delta_1}.
\end{equation}
As a result, we can derive that
\begin{equation}
\begin{aligned}
\|\tilde{d}_v\|^2&=\|\tilde{d}_{v\parallel}\|^2+\|\tilde{d}_{v\perp}\|^2\\
&\le\|\epsilon\|^2(\frac{c_0^2}{\delta_1}+\frac{1}{\lambda_1\lambda_{\min}(L)})\\
\|\tilde{d}_u\|^2&=\|\tilde{d}_{u\parallel}\|^2+\|\tilde{d}_{u\perp}\|^2\\
&\le\|\epsilon\|^2(\frac{c_0^2}{\delta_1}+\frac{1}{\lambda_2\lambda_{\min}(H)}).
\end{aligned}
\end{equation}
Considering that $\tilde{d}_v = d_v + \epsilon_1$ and $\tilde{d}_u = d_u + \epsilon_y$, we can bound the error $d_v$ and $d_u$:
\begin{equation}
\begin{aligned}
\|d_v\|^2 &\le \|\tilde{d}_v\|^2 + \|\epsilon_1\|^2 \\
&\le \|\epsilon\|^2(\frac{c_0^2}{\delta_1}+\frac{1}{\lambda_2\lambda_{\min}(H)})+\|\epsilon_1\|^2\\
&=\tilde{c_1}\|\epsilon\|^2+\|\epsilon_1\|^2\\
\|d_u\|^2 &\le \|\tilde{d}_u\|^2 + \|\epsilon_2\|^2 \\
&\le \|\epsilon\|^2(\frac{c_0^2}{\delta_1}+\frac{1}{\lambda_1\lambda_{\min}(L)})+\|\epsilon_2\|^2\\
&=\tilde{c_2}\|\epsilon\|^2+\|\epsilon_2\|^2.
\end{aligned}
\end{equation}
Noticing that 
\begin{equation}
\begin{aligned}
\tilde{c_1}&\le \frac{2}{\delta_1}(2+\frac{1}{\sqrt{\lambda_1\lambda_{\min}(L)}}+\frac{1}{\sqrt{\lambda_2\lambda_{\min}(H)}})^2\\
\tilde{c_2}&\le \frac{2}{\delta_1}(2+\frac{1}{\sqrt{\lambda_1\lambda_{\min}(L)}}+\frac{1}{\sqrt{\lambda_2\lambda_{\min}(H)}})^2,
\end{aligned}
\end{equation}
and that
\begin{equation}
\begin{aligned}
\epsilon_v &= \epsilon_1^TL\epsilon_1\ge \lambda_{\min}(L)\|\epsilon_1\|^2\\
\epsilon_u &= \epsilon_2^TH\epsilon_2\ge \lambda_{\min}(H)\|\epsilon_2\|^2.
\end{aligned}
\end{equation}
we achieve the bound in Theorem \ref{ns}.
\subsection{Proof of Proposition \ref{nsl}}
The result can be derived from the proof of Theorem \ref{ns}.

\end{document}